# Inconsistencies of metalens performance and comparison with conventional diffractive optics

**Rajesh Menon**[1,2,*] **and Berardi Sensale-Rodriguez**[1]

[1]Department of Electrical & Computer Engineering, University of Utah, 50 Central Campus Dr. Salt Lake City UT 84112

[2]Oblate Optics, Inc. San Diego CA 92130

[*]rmenon@eng.utah.edu

**Abstract:** We posit that inconsistent interpretations of experimental data have led to inaccurate claims on metalens focusing efficiencies. By performing a meta-analysis, we show that extraordinary claims of high focusing efficiency at high numerical apertures are, unfortunately, not yet backed by rigorous simulation or experimental results.

The Fresnel-zone plate (FZP), as well as their generalizations via multilevel and effective-medium implementations, including those containing sub-wavelength features, have been traditionally considered under the umbrella of diffractive lenses [1]. Recently, flat lenses containing subwavelength features have been termed "metalenses" [2]. This definition encompasses phase FZPs with numerical aperture (NA) > 0.5, as well as related devices like photon sieves, superoscillating flat lenses, super-lenses, transformation-optics-based lenses, radially-polarized focusing lenses, photonics-crystal-based lenses, and others. It is noted that effective-medium theory was used to map desired phase transformation into subwavelength features before the introduction of the term "metalens" [3]. The term "metalens" is quite comprehensive, making it somewhat unnecessary to differentiate it from diffractive lenses. However, proponents of metalenses claim two advantages: higher focusing efficiencies at high NA and multi-functionality, suggesting that diffractive lenses lack these attributes [2,4]. Unfortunately, we find that there is no clear evidence to justify either of these claims. This comment urges the scientific community to conduct systematic research to substantiate these benefits.

The focusing efficiency of diffractive lenses decreases at high NA [5]. But, this decrease can be avoided by local-geometry optimization, [6] without resorting to any new physics.

An analysis of the published literature on metalenses shows that many reports of metalens-focusing efficiencies might be inaccurate or inconsistent. The focusing efficiency, $\eta$ is the fraction of incident power that is focused. The focal power is typically measured with an iris, whose radius ranges from $3\times$ to $\geq 18\times$ the full-width-at-half-maximum (FWHM) of the focal spot, and sometimes with no iris at all [7]. If all incident light is diffracted into one converging spherical wave and when NA $\lessapprox 0.8$, then the iris radius is not important as long as it is greater than $\sim 3\times$ FWHM (see the ideal Airy disk in Fig. 1a). However, a real flat lens diffracts light into multiple orders (see Fig. 1b). Then, a large iris radius grossly over-estimates focusing efficiency. For example, refs.[8] and [9] use irises of radius $8.5\times$, $40\times$ FWHM, respectively, while ref. [7] employs no iris. In some cases, focusing efficiencies higher than those theoretically possible (bounded by a unit-cell-based design) [10] have been reported (see Table S1 in the supplement). The unit-cell-based design is not capable of precisely representing the necessary phase transformation, which results in undesired diffraction orders. This problem was thoroughly analyzed several decades ago, [11] and more recently revisited [12]. To build upon [12], here we have quantitatively analyzed the details from numerous papers and clarified where these claims might be invalid (see supplement). Although there are discrepancies in the proposed characterization methods in [12] (see supplement), we agree with the key message that it is important to establish a direct comparison between the focusing efficiency (and performance, in general) of a metalens and that of an optimized diffractive lens (similar to the ones described in ref. [6]) with the same specifications.

There might also be inaccuracies in the calculations of the Strehl Ratio (SR) and the modulation-transfer function (MTF). To obtain SR, the measured PSF is normalized to the Airy function by matching their encircled powers (area under the curve of the PSF). For example, in ref. [13], this normalization is performed by integrating the power over a radius of only $3\times$ FWHM. The MTF is often calculated as the absolute value of the Fourier transform of the PSF. However, the PSF is often prematurely truncated (often due to the limited field-of-view of a magnifying objective). This causes both the SR and MTF to be over-estimated.

The second advantage of multi-functionality is claimed based on the observation that a metalens "can function differently depending on different degrees of freedom of light (wavelength, polarization, incident angle and so on)" [4]. We note that "conventional" diffractive optics have long performed different functions based on polarization, [14] wavelength, [15] and incident angle. [16]

By dramatically increasing the number of degrees of freedom, sub-wavelength diffractive optics offer a treasure trove of possibilities for imaging and inferencing. However, there is a clear need to establish these advantages with rigorous experiments in order to advance their adoption.

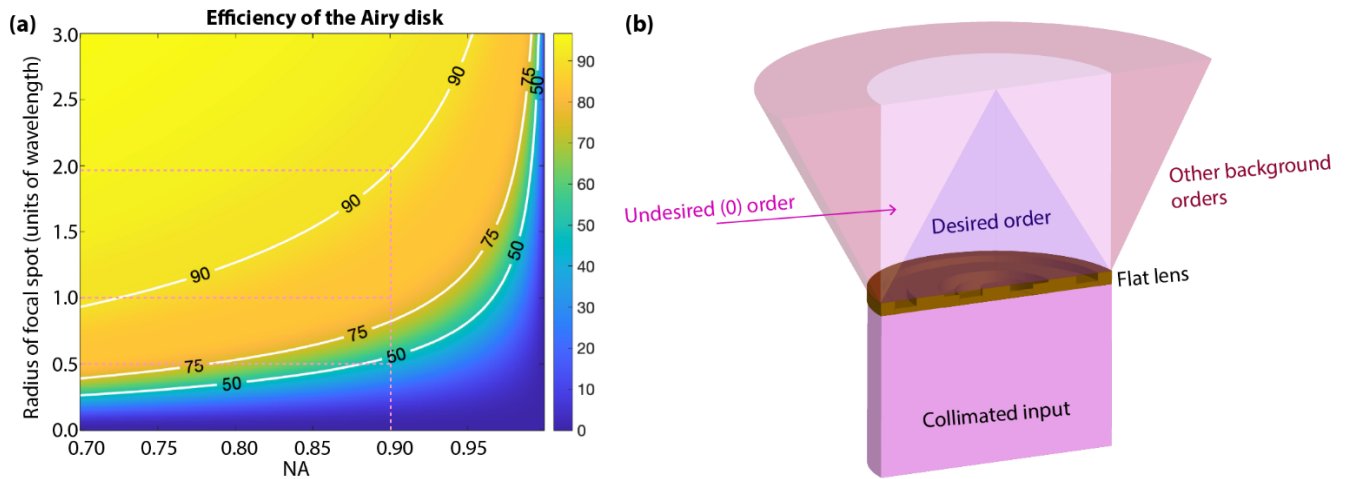

**Figure 1. Focusing efficiency of the ideal Airy disk and illustration of diffraction orders from flat lenses.** **(a)** Focusing efficiency, $\eta$ of the ideal Airy disk as function of NA and the radius of the focal spot, R (normalized to FWHM=$0.5\lambda$/NA). For NA $\lesssim 0.8$, $\eta$ is stable for R > 3×FWHM. But for higher NA, $\eta$ varies rapidly with R. **(b)** All real flat lenses generate not only the desired diffraction order, but a multitude of undesired orders. Therefore, if R≫FWHM, then focusing efficiency can be grossly over-estimated.

## Acknowledgements

Support from grants #N6560-NV-ONR and NASA #NNL16AA05C are gratefully acknowledged. The authors thank H. Smith, H. Li and O. Miller for fruitful discussions.

## Author contributions statement

All authors reviewed the manuscript.

**Competing interests** RM has financial interest in Oblate Optics, Inc., which is commercializing flat lenses.

## 1. Diffraction orders:

When a diffraction grating is illuminated by a plane wave, the transmitted field is often expressed as a linear combination of plane waves, referred to as diffraction orders. The ratio of the power carried away by each plane wave to the total incident power is referred to as the diffraction efficiency. Although less well known, this analysis can be applied to diffractive lenses as well [1]. As illustrated below, the transmitted field can be easily decomposed into a linear combination of spherical waves (converging and diverging), which form a complete basis set. In the case of diffractive lenses, the diffraction efficiency is similarly defined as the ratio of power carried by one of these spherical waves to the total incident power. The convention of notation for these orders is illustrated in Fig. 1(a).

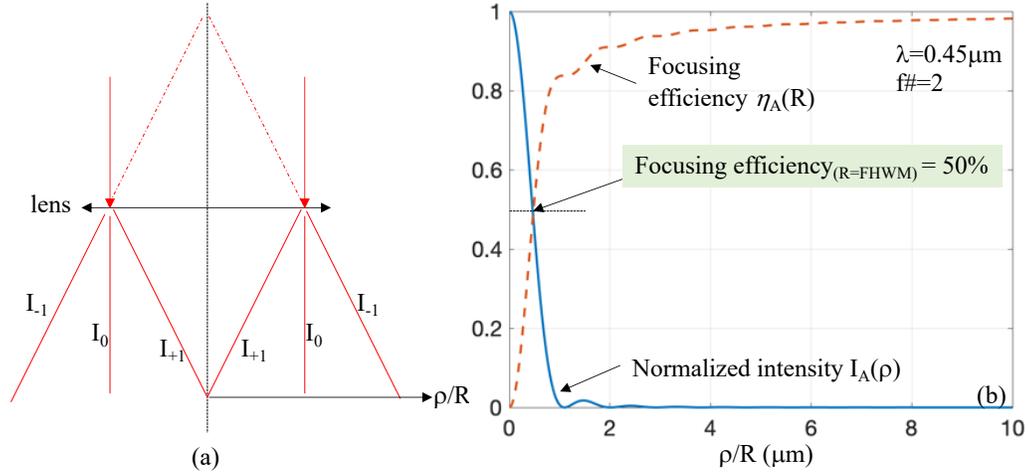

*Figure S1: (a) Diffraction orders of a lens are the spherical wave decomposition of the transmitted field. (b) Airy function is the PSF of a lens with 100% diffraction efficiency into the 1st order. Focusing efficiency is defined as the ratio of power inside a spot of radius R in the focal plane to the incident power. This is also referred to as the encircled power. In the case of an ideal lens (Airy function), it is equal to 50% when the radius R is equal to the full-width at half-maximum of the PSF.*

Consequently, the point-spread function (PSF), which is the intensity distribution in the focal plane of a lens when the lens is illuminated by a plane wave can be expressed as: $I(\rho) = I_0 + I_{+1}(\rho) + I_{-1}(\rho) + ...$, where $I_0$ is the spatially uniform 0th diffracted order, the subscript refers to the order of diffraction, the sign in the subscript refers to whether the spherical wave is converging (+) or diverging (-), and the sum occurs for all propagating diffracted orders. Note that we assumed the simple case of cylindrical symmetry (and normal incidence or on-axis focusing) for simplicity, but the principle applies generally.

The encircled power is defined as $E(R) = \int_0^R I(\rho)\rho d\rho$, where R is the radius of the spot over which the power is encircled. If one ignores Fresnel reflection losses at the interfaces and absorption losses in the material, then $E(R = \infty) = E_0$, the total incident power. Then, the focusing efficiency is given simply by $\eta(R) = \frac{E(R)}{E_0}$. As expected, the focusing efficiency is dependent upon the size of the focal spot, R. In the case of polarization sensitive lenses, it is important to ensure that $E_0$ accounts for incident power with unpolarized light (if this metric is used to compare to conventional polarization insensitive lenses).

Next, one may ask the question what is the focusing efficiency of an ideal lens. Diffraction ensures that a lens cannot have 100% focusing efficiency, even when the diffraction efficiency of one order may be 100%. To illustrate this point, we can consider an ideal lens with 100% diffraction efficiency into its 1$^{st}$ order. The PSF of such a lens is simply the Airy function, $I_A(\rho) = I_{A0}\left(\frac{2J_1(x)}{x}\right)^2$, where $I_{A0}$ is the incident intensity, $J_1(x)$ is the Bessel function of the 1$^{st}$ kind, and $x = \frac{\pi\rho}{\lambda f\#}$, where λ is the wavelength and $f\#$ is the f number of the lens. The encircled power of the Airy function can be analytically expressed as: $E_A(R) = \int_0^R I_A(\rho)\rho d\rho = E_{A0}\left(1 - J_0^2\left(\frac{\pi R}{\lambda f\#}\right) - J_1^2\left(\frac{\pi R}{\lambda f\#}\right)\right)$, where $E_{A0}$ is the total incident power. Then, the focusing efficiency is simply given by: $\eta_A(R) = \left(1 - J_0^2\left(\frac{\pi R}{\lambda f\#}\right) - J_1^2\left(\frac{\pi R}{\lambda f\#}\right)\right)$. From this expression, we can readily show that the focusing efficiency when $R$ = full-width at half-maximum (FWHM) of the PSF is 50% (see Fig. S2b). In other words, a lens with 100% diffraction efficiency can have focusing efficiency of 50% (measured with $R$=FWHM). We further emphasize that reporting focusing efficiency without reporting the size of the focused spot is incomplete. Complete information about focusing efficiency requires the encircled power.

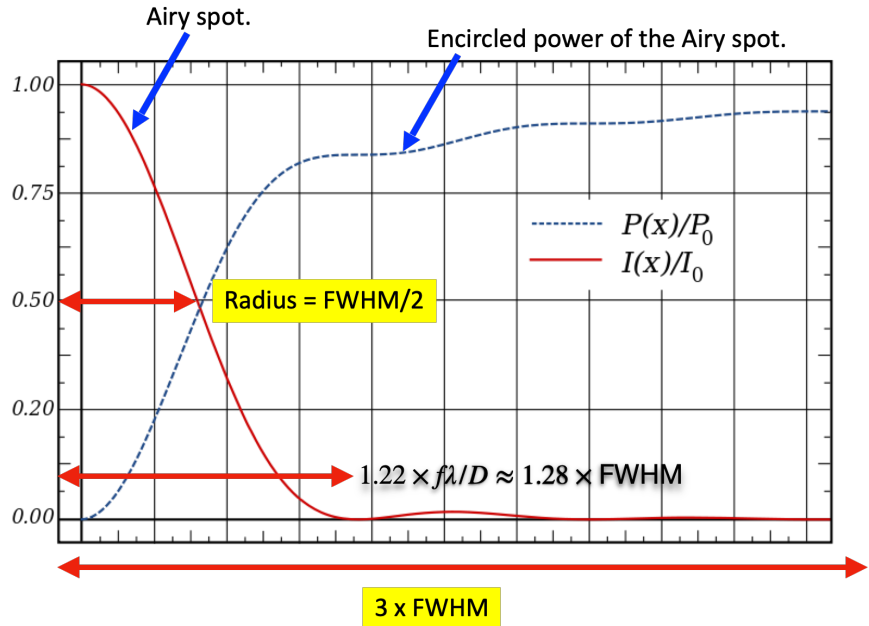

*Figure S2: Airy spot and its encircled power.*

## 2. Notes on efficiency reports:

*Table S1: Summary of reported metalens efficiencies (only one example from each reference is shown for brevity). The bound from ref [2] is computed based on the unit-cell period and NA from Fig. 2a in [2]. All values in this table are for monochromatic lenses. Those values marked in red exceed bounds dictated by the unit-cell design method [2]. {\*focal spot size = 5 X FWHM. \*\* focal spot size = 4 X FWHM.}*

| Reference | λ | NA | Unit-cell period (Λ/λ) | Radius of focal spot (/FWHM) | Reported eff. | Bound from Ref [2] |
|---|---|---|---|---|---|---|
| [4] | 405nm | 0.8 | 0.49 | Not disclosed | 88% | 64.6%* |
| [6] | 660nm | 0.85 | 0.53 | 2 sim, 4 expt. | 84% sim, 60% expt. | 56.5%** |
| [7] | 532nm | 0.9 | 0.45 | Not disclosed | 42% | 58.5%* |
| [8] | 532nm | 0.98 | 0.41 | 3 | 67% | 45% |
| [9] | 405nm | 0.8 | 0.49 | Not disclosed | 86% | 64.6%* |
| [10] | 915nm | 0.95 | Grating averaging | 10 | 77% | - |
| [11] | 1,550nm | 0.89 | 0.52 | 3 | 72% | 51.9% |

In Ref [4], the supplement includes this statement:
"*For efficiency measurements, we used a supercontinuum laser (SuperK) as the source. The efficiency is defined as the ratio of the optical power of the focused beam to the optical power of the incident beam, as captured by a photodetector (Thorlabs S120C) located at the same position as the CMOS camera. The incident optical power was measured as the light passing through an aperture (aluminum on glass) with the same size as the metalens.*"
Based on this text and the Fig. S2 in the supplement of ref [4], we surmise that an aperture (or iris) was likely not used to measure the focused power. Therefore, this reported efficiency may include both the desired and undesired diffracted orders.

In Ref. [5], Fig. S3 in the supplement indicates that there is no aperture in the focal plane when measuring the focused power. Therefore, we speculate that the reported efficiency might have the same issue as in Ref. [4].

Ref [6] states:
"*The efficiency is calculated as the ratio of the optical power of the measured focused beam to that of the incident beam. The incident beam was measured as the optical power passing through a circular aperture (aluminum on glass) with the same diameter (300 μm) as the metalenses.*"
Later with regard to Fig 4, it states:
"*We define the efficiency as the ratio of the optical power in the focal spot area (circle of radius 2 × FWM spanning the center of the focal spot) to the incident optical power.*"
We further note that the experimental efficiencies are from results in Fig. S3b, while simulations are shown in Fig. 4. In the caption of Fig. S2b (regarding measurement of focusing efficiencies), it states:
"*The diameter of the iris was about eight times of the FWHMs of the focal spots of the metalenses (at design wavelength) on the image plane of the set-up.*"
So, the radius of focal spot is different for simulation and experiment. We computed the bound for the larger radius (4 x FWHM) to be conservative.

In Ref [7], the main text states:
"*The focusing efficiency is defined as the power of focal spot divided by incident light in the case of circularly polarized light.*"
and the caption of Fig. S3 says:
"*The efficiency was measured, in case of circularly polarized incidence, by dividing the power of the focal spot by the total power passing through an aperture with the same diameter as the water immersion meta-lens.*"
Otherwise, no information is provided regarding the radius of the focal spot used for measuring efficiencies. We used a radius of 5 X FWHM to compute the bound.

Ref [8] states:
"*the focusing efficiency can be defined as the fraction of the incident light that passes through a circular iris in the focal plane with a radius equal to three times the fwhm spot size, so defined as the total power in the desired focal spot divided by the total incident power.*"

Ref [9] states:
*"For efficiency measurements, a tunable laser (SuperK, NTK Photonics) was used and the CMOS camera was replaced by a photodetector (Thorlabs S120C). The efficiency is defined as the ratio of the optical power of the focused beam to the optical power of the incident beam. The latter was measured as the optical power passing through a circular aperture (aluminum on glass) with the same diameter as the metalenses.*"
There is no other information about the radius of the focal spot.

Ref [10] says:
"*The focusing efficiency is defined as the percentage of the power incident on the metalens aperture that is focused into and passes through a circle with a radius of 5 μm centered around the focal spot of the metalens. The radius of 5 μm for the aperture is selected for a direct comparison with the experimentally measured results that are presented in the next section.*"
For the experiments, it says:
"*We measured the focusing efficiency of the metalens using the setup shown in Fig. 4f.*"
"*The intensity distribution at the metalens focal plane was magnified by 100×, masked by passing through a 1-mm-diameter aperture in the image plane (corresponding to a 5-μm-radius aperture in the metalens focal plane), and its power was measured. To measure the incident optical power, we focused the incident beam using a commercial lens (Thorlabs AC254- 030-B-ML with a focal length of 3 cm and a transmission efficiency of 98%) and measured its power in the image plane (Fig. 4g). The focusing efficiency of the metalens was found as 77% by dividing the power passed through the aperture in Fig. 4f to the incident power.*"
From Fig. 4(d), we can estimate the FWHM as 0.5μm, which gives a spot radius of 10 X FWHM.

Ref [11] says:
"*we define the focusing efficiency as the fraction of the incident light that passes through a circular aperture in the plane of focus with a radius equal to three times the FWHM spot size.*'

## 3. Notes on ref [12]

In ref. [12], the authors make a similar argument as in our comment, but with the emphasis on a need for standardizing the characterization of flat lenses. However, there are a few incorrect points made that we would like to note here.

It is suggested that the diffraction efficiency is the correct metric for flat lenses and the authors state: "When we measure the power going to the design order, we need to collect light over the same area over which we measured the PSF, that is, until the PSF flattens out." Unfortunately, this last phrase is problematic as the PSF might flatten out with a large dc background, which would correspond to a large fraction of incident power going to the zero order. The subsequent comments about the use of this PSF is compromised if there is power in the zero (undiffracted) order. By the way, this issue has been thoroughly studied in Ref. [13].

A second problem is actually acknowledged by the authors themselves: "Therefore, the stray light, which is spread over a very large area, cannot be included in the MTF measurement (Fig. 1a)." This conveniently ignores not only the zero order, but any additional orders in the MTF. Therefore, comparing this "artificially enhanced" MTF to the diffraction-limited MTF (which is what the authors suggest) is clearly not accurate.

Notwithstanding these issues, we are in agreement with the authors of this reference, in particular their statement: "The first and perhaps the most basic problem is not comparing achromatic flat lens performance to that of a conventional flat lens."

**Supplement References:**

# Expanded references to main text discussion:

In the main text of our comment, we refer to several types of flat optics containing subwavelength features, following below are example references of these: photon sieves [S1], superoscillating flat lenses [S2, S3], super-lenses [S4], transformation-optics-based lenses [S5, S6], radially-polarized focusing lens [S7], photonics-crystal-based lenses [S8]. Furthermore, it is to mention that sub-wavelength features have been applied in diffractive lenses directly to map the required phase transformation [S9], using effective-medium theory (and sub-wavelength dielectric pillars) [S10, S11], and also with optimization-based inverse design [S12, S13].

It is also mentioned in the main text that focusing efficiency of high-NA multi-level diffractive lenses was calculated using rigorous electromagnetic theory in a seminal paper published in 2001 [S14]. However, it is to be noted that even before this work, it was shown that the sharp drop-off in efficiency for high-NA diffractive lenses can be avoided by optimization of the local geometry, both in simulations [S15, S16] and in experiments [S17]. Furthermore, it was shown that even higher efficiencies could be obtained in the context of imaging at high NA [S18].

As discussed, if the PSF is prematurely truncated, the MTF can be grossly overestimated. It is to note that these errors lead to reports of a high Strehl ratio with modest focusing efficiencies [S19, S20], which is not mathematically possible.

Here we also provide more references to multi-functionality of diffractive optics: based on polarization [S21], wavelength [S22-S24] and incident angle [S25].

.

**Email confirming date of submission of Errata as March 23, 2024.**

Hi David,

Hope you are well. I'm writing to submit an errata to the subject comment. Please find the corresponding documents attached. Can you please advise what the correct procedure for submitting this is.

Many thanks,
Rajesh Menon.

**Nat_Pho_Comment-Correction_v5.docx**

**supplemental-document-v5.docx**

# Author correction: Inconsistencies of metalens performance and comparison with conventional diffractive optics

**Rajesh Menon**[1,2,*] **and Berardi Sensale-Rodriguez**[1]

[1]Department of Electrical & Computer Engineering, University of Utah, 50 Central Campus Dr. Salt Lake City UT 84112

[2]Oblate Optics, Inc. San Diego CA 92130

*rmenon@eng.utah.edu

In the version of this Comment initially published, the plot in **Fig. 1a** was generated with an error in the equation that converted numerical aperture (NA) to f-number. Furthermore, we were made aware of the fact that the Airy function is not a good representative of the field distribution in the focal plane of a high-NA idealized flat lens. We therefore employed the formulation described in Ref [1] to replot this figure. The corrected plot shows similar trends to the original plot, supporting the case that we made originally, which is that the choice of the radius of the iris used to measure the power in the focal spot is important when computing efficiency.

It was also brought to our attention by the authors of Ref. [2] of a mistake in that paper, which is currently being corrected by an errata that those authors are preparing. As a result, we would like to replace our citation to Ref. [10] in the published comment (which is Ref. [2] below) with Ref. [1] below. The Ref. [1] serves the same purpose. In other words, the upper limit of efficiency achievable by an ideal metalens is computed for various high NA lenses in Ref. [1]. Thereby, we also updated Table S1 in our supplement. Note that in Table S1 we incorrectly reported the NA of Ref. [11] from the supplement as 0.95, when it was actually 0.78. This has also been corrected.

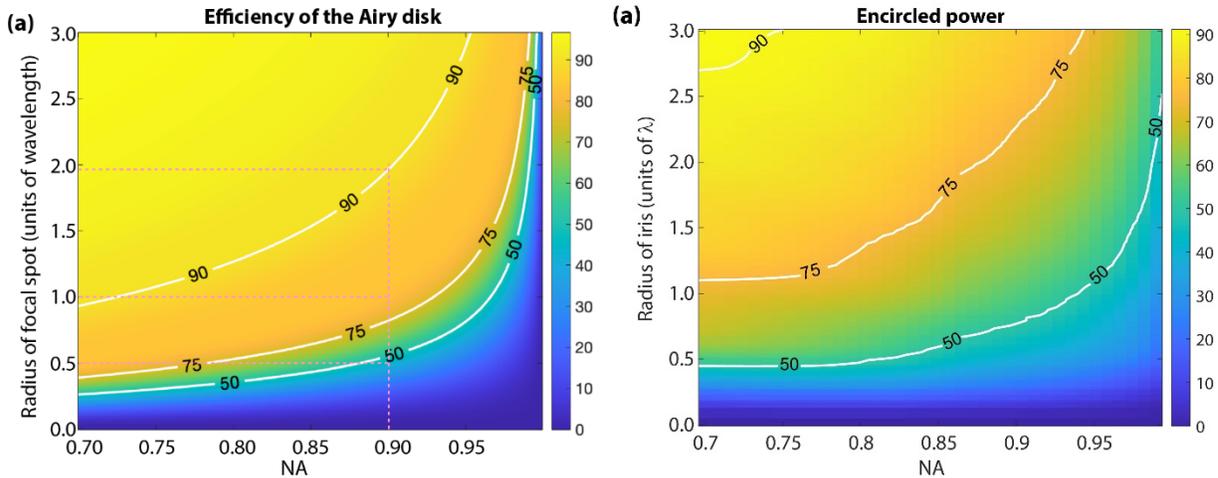

**Figure 1a.** Original (left) and corrected (right). Corrected caption should read: Encircled power by an ideal metalens as function of NA and the radius of the iris in units of λ, where λ is wavelength.

Lastly, we reiterate the point of our comment that, to our knowledge, there is no clear evidence showing the advantage of metalenses over conventional diffractive lenses at high NA (e.g., NA > ~0.9). We believe a useful approach for all experimenters is to report the results of metalenses along with a conventional diffractive lens (Fresnel zone plate) of the same specifications.

# Supplementary Information for Inconsistencies of metalens performance and comparison with conventional diffractive optics.

*Revised March 21, 2024 to correct errors from prior version.*

Rajesh Menon[1,2,*] & Berardi Sensale-Rodriguez

[1]Department of Electrical & Computer Engineering, University of Utah, 50 Central Campus Dr., Salt Lake City, UT 84112, USA.

[2]Oblate Optics, Inc., San Diego, CA 92130, USA

[*]rmenon@eng.utah.edu

Table S1: Summary of reported metalens efficiencies (only one example from each reference is shown for brevity). All values in this table are for monochromatic lenses. Those values marked in red exceed bounds dictated by Ref. [12].

| Reference | λ | NA | Unit-cell period (Λ/λ) | Radius of focal spot (/FWHM) | Reported eff. | Eff from Fig. S4 (Radius of 3 x FWHM) |
|---|---|---|---|---|---|---|
| [4] | 405nm | 0.8 | 0.49 | Not disclosed | 88% | 83% |
| [6] | 660nm | 0.85 | 0.53 | 2 sim, 4 expt. | 84% sim, 60% expt. | 79% |
| [7] | 532nm | 0.9 | 0.45 | Not disclosed | 42% | 74% |
| [8] | 532nm | 0.98 | 0.41 | 3 | 67% | 49% |
| [9] | 405nm | 0.8 | 0.49 | Not disclosed | 86% | 83% |
| [10] | 915nm | 0.78 | Grating averaging | 10 | 77% | 84% |
| [11] | 1,550nm | 0.89 | 0.52 | 3 | 72% | 75% |

**Email responding to comment dated June 21, 2024.**

Hi David,

Thank you for the note and sharing with us the new comment. We have been in touch with the authors and have had many discussions with them, some of which precipitated our Erratum. Here is a brief summary of our responses to this new comment.

Sincerely,
Rajesh & Berardi.

First, we note that the main critiques of the comment have been addressed in our errata submitted on March 23, 2024. Our revised figure (reproduced below) is consistent with Fig. 1d in the new comment. These simulations support our main point that focusing efficiency (and more precisely, encircled power) is sensitive to the choice of aperture radius, especially at high NA. This sensitivity is highlighted by the increasing slope of the contours in the plot with respect to NA. Therefore, we urge you to publish our erratum as soon as possible.

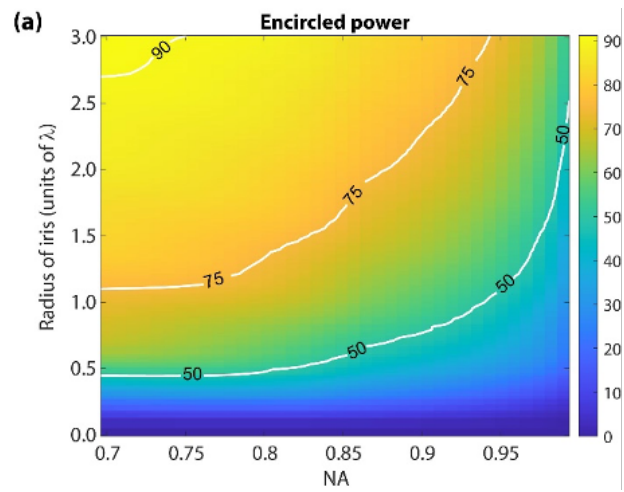

Next, we emphasize that the current comment implicitly supports our point. The authors state: "The fraction of the incident power that is directed toward the main focal point is referred to as the (absolute) focusing efficiency and is used to compare the performance of different flat lens designs." This definition is ambiguous, and this is an important point that our original comment aimed to bring up to the community: what does "main focal point" mean? Different publications resolve this ambiguity differently, defining the main focal point with varying spot radii, such as 3 times the full-width-at-half-maximum (FWHM), 6 times FWHM, or even much larger.

The authors also state: "We refer to a local flat lens with 100% focusing efficiency that is corrected for spherical aberration as an ideal flat lens." This statement is inconsistent with the previously quoted comment, especially when the numerical aperture (NA) is high ($> \sim 0.7$). With such a definition (which as we note is inconsistent), achieving 100% focusing efficiency requires a spot radius much larger than the FWHM at high NA, which is impractical in microscopy (perhaps the most important application of such lenses). High-NA lenses are used to achieve high lateral resolution, requiring sufficient power within a small spot for a high signal-to-noise ratio. If efficiency is not defined relative to the wavelength (or FWHM), it is not a useful metric. This inconsistency in defining focusing efficiency across papers is the main point of our original comment, which remains valid.

Next, we highlight this statement: "Compared to multi-functional devices implemented using multi-level diffractive optics, metalenses offer **higher efficiency**, higher spatial resolution for sampling wavefronts, and simpler fabrication." (bolded by us). Claims that metalenses offer higher efficiency than conventional diffractive optics are often made without citation or experimental evidence, as above, again another important point that we make in our comment.

One key point of our comment is that, until recently, no comparison existed between a metalens and a conventional diffractive lens with the same specifications to support this claim. No citation for this claim is provided in the comment or in Ref. 2 cited in the comment. Recently, we became aware of a paper that performed this comparison both experimentally and through simulations: arXiv:2401.13427. This work (see Fig. 4) shows that the performance of the metalens and the Fresnel zone plate (FZP, a conventional diffractive lens) are very similar. In fact, the metalens performed slightly worse, with a simulated efficiency of 61% compared to 68% for the FZP.

The same point applies to the last critique on multi-functionality. As far as we know, no comparison has been made between multi-functional diffractive optics and multi-functional metalenses with the same functionalities to support the claim made in this comment and in Ref. 2 (and many other publications). Our point is to remind readers, especially the next generation of scientists, that multi-functional diffractive optics have existed for many decades. Claiming superiority over these requires rigorous experimental data, and an apples-to-apples comparison, which is unfortunately not yet available. While we are optimistic that new capabilities will emerge, the evidence is not there yet.

> On Mar 23, 2024, at 6:00 AM, RAJESH MENON <rmenon@eng.utah.edu> wrote:
>
> Hi David,
>
> Hope you are well. I'm writing to submit an errata to the subject comment. Please find the corresponding documents attached. Can you please advise what the correct procedure for submitting this is.
>
> Many thanks,
> Rajesh Menon.
>
> <Nat_Pho_Comment-Correction_v5.docx><supplemental-document-v5.docx>